\documentclass[aps,preprint,preprintnumbers,nofootinbib,showpacs]{revtex4}
\usepackage{epsfig}
\usepackage{color}
\usepackage{amsmath}
\usepackage{amssymb}
\usepackage{dsfont}

\def\bea{\begin{eqnarray}}
\def\eea{\end{eqnarray}}
\def\sea{\nonumber \\&&}
\def\lla{\left\langle}
\def\rra{\right\rangle}

\def\ssc{\scriptscriptstyle}
\def\lsim{\mathrel{\raise.3ex\hbox{$<$\kern-.75em\lower1ex\hbox{$\sim$}}} }
\def\gsim{\mathrel{\raise.3ex\hbox{$>$\kern-.75em\lower1ex\hbox{$\sim$}}} }

\begin{document}

\draft
\preprint{{\vbox{\hbox{NCU-HEP-k092}
\hbox{Dec 2021}
\hbox{rev. Mar 2022}
\hbox{ed. Jul 2022}
}}}
\vspace*{.5in}

\title{Mixed State Parametrization and Two-qubit Entanglement

\vspace*{.3in} }
\author{Otto C. W. Kong$^*$ and Hock King Ting}

\address{Department of Physics and 
Center for High Energy and High Field Physics,\\
National Central University, Chung-Li 32054, Taiwan \\
$^*$Corresponding author. E-mail : otto@phy.ncu.edu.tw 
\vspace*{1.5in}}

\begin{abstract}\vspace*{.3in}
A generic scheme for the parametrization of mixed state systems
is introduced. A slightly modified  version specially for bipartite 
systems is then given, and especially applied to a two-qubit system. 
Various features of two-qubit entanglement are analyzed based on 
the scheme. Our  formulation of the parametrization and the analysis 
of entanglement properties exploit the interplay between pure states 
as Hilbert space vectors and pure as well as mixed states as density 
matrices. Explicit entanglement results are presented, in terms of 
negativity and concurrence, for all two-qubit mixed states with one 
single parameter/coordinate among the full set of fifteen being 
zero and a few other interesting cases.
\end{abstract}

\maketitle

\section{Introduction}
Entanglement, as the characteristic feature of quantum physics
and a great resource for quantum information technology, is at 
the heart of modern interest in physical science and technology
 \cite{BZ,NC,H4}.  However, more than thirty years after the paper
by Werner \cite{We}, which effectively kick-started a very active
period of studies in various aspects of quantum entanglement,
especially for mixed states, the `simple' notion of a complete
characterization of the simplest system with entanglement,
namely a system of two-qubits, has not been available. We still
do not have a general expression that gives the separability
condition of a generic two-qubit mixed state in a single parametric 
form, not to say a general expression for even one measure of
entanglement for any such inseparable states. 

We present here an approach to the complete parametrization 
of any pure or mixed states of any finite-dimensional quantum 
system, which is an extension of that for the pure states to general 
mixed states, each with the pure state as the one among the class 
that is closest to the mixed state. It is then applied, with a small 
adjustment, to the case of bipartite systems, with the dominating 
pure state part as a generally entangled extension of that of
a product state which is again closest to it.

As an alternative to parametrizations of mixed states or 
coordinations of the geometric space of all of them for 
a quantum system \cite{p1,BZ}, our approach is marked by 
using strongly the links between pure states as normalized
vectors in the corresponding Hilbert spaces and mixed 
states as their eigenensemble of pure states from their 
purification as pure states of a doubled system. It has 
advantages in connection with those links. In particular, we 
illustrate that through the application to the two-qubit case, for 
which we get entanglement results of all mixed states with 
one single parameter/coordinate among the full set of 
fifteen being zero. Results for a few other limiting cases 
are also presented. We have interesting results on explicit 
separability conditions and entanglement measures 
apparently not available in the literature. Our scheme,
however, is more than just an alternative. It can be easily
seen that the scheme incorporates some physically 
important parameters into the full coordination of the
space of mixed states, hence is also wells suited for
analyzing the geometric structure of the latter and its
relationship with the entanglement features.

Apart from individual results, we illustrate some interesting features
on how the entanglements of parts of a pure or mixed state contribute 
to the overall entanglement, with interference of entanglements in 
the pure states and (partial) cancellation of entanglements in the 
mixed states. The last has been generally appreciated in the notion 
of robustness of entanglement \cite{VT}. As the positive partial
transpose (PPT) condition is necessary and sufficient for separability 
of two-qubit states \cite{P,H}, we use the negativity \cite{Z+,VW}, as 
a natural measure of PPT violation, as well as the perhaps more 
popular concurrence \cite{W,V+}  to study the entanglement. The 
two measures have interesting similarities and differences in their 
dependence on the basic parameters. The results are expected to 
be illustrating features that are in a way generic for systems with 
entanglement. They, in a way, complement the studies in 
Refs.\cite{V+,MG}. 

\section{An Approach to Parametrization of Mixed States}
\subsection{A Generic Parametrization of Mixed States for a Quantum System}
The general expression for a single qubit mixed state 
can be written as
\bea
\frac{1+r}{2} \left|\phi\rra\!\!\lla \phi\right|
              + \frac{1-r}{2}  \left|\phi_\perp\rra\!\!\lla \phi_\perp\right|
&=& \! \left( \frac{1+r}{2}|c|^2 +\frac{1-r}{2}|s|^2 \right) \! \left|0\rra\!\!\lla 0\right|
\sea \hspace*{-1in}
  + \! \left( \frac{1+r}{2}|s|^2 +\frac{1-r}{2}|c|^2 \right)  \!  \left|1\rra\!\!\lla 1\right|
 + r \bar{s} {c} \left|0\rra\!\!\lla 1\right| 
  + r \bar{c} {s} \left|1\rra\!\!\lla 0\right|, \!
\eea
where $ \left|\phi\rra= {c}   \left|0\rra + {s}  \left|1\rra$,
$ \left|\phi_\perp\rra= -\bar{s}   \left|0\rra + \bar{c}  \left|1\rra$,
${c}\equiv\cos(\frac{\theta}{2}) e^{\frac{-i\psi}{2}}$ and
${s}\equiv \sin(\frac{\theta}{2}) e^{\frac{i\psi}{2}}$,
$0 \leq \theta < \pi$, $0 \leq \psi < 2\pi$. 
with $0\leq r\leq 1$ a measure of the purity of the state.
The generic pure state  $ \left|\phi\rra$ is  
given by a unitary transformation acting on the fixed state
 $\left| 0\rra$, as $U\left| 0\rra$, taking the $U$ as an
operator, the matrix form of which is given by
$\left( \begin{array}{cc}
c & - \bar{s} \\
s & \bar{c} \\
\end{array}  \right)$. 
And $ \left|\phi_\perp\rra = U\left| 1\rra$; all state vectors 
normalized. It is important to note that while the full symmetry 
on the Hilbert space is an $SU(2)$, which is a three-parameter 
group, $U$ lives only in a subgroup of two parameters. There 
is a nontrivial little group for $\left| 0\rra$ which does not 
change the vector. We elaborate on that because the structure 
is what we will use repeatedly below to get the general 
parametrization, and something that has been overlooked 
in some of the literature.

We can write a generic pure two-qubit state as 
$\left| \Psi  \rra = q_+ \left| \phi \phi' \rra + e^{i\zeta} q_- \left| \phi_\perp \phi'_\perp \rra$,
or explicitly with the six real independent parameters as
\bea&&
 q_+ (\,{c} {c}' \left| 00 \rra + {c}{s}'  \left| 01 \rra 
  + {s} {c}' \left| 10 \rra + {s}{s}'  \left| 11 \rra  )
  + e^{i\zeta} q_- (\, \bar{s} \bar{s}'  \left| 00 \rra - \bar{s} \bar{c}'
    \left| 01 \rra  - \bar{c} \bar{s}' \left| 10 \rra +  \bar{c} \bar{c}'  \left| 11 \rra ) \;,
\sea
\label{2qps}\eea
where $q_\pm=\sqrt{\frac{1\pm r}{2}}$. The reduced density matrix 
$\rho_{\!\ssc A}$  is exactly given by our expression of the general 
mixed state density matrix  for the single qubit system; $\rho_{\!\ssc B}$
is of the same form with ${c}$ and ${s}$ replaced 
by ${c}'$ and ${s}'$. 
We note that the concurrence \cite{W}, as the essentially unique
measure of the entanglement of the state, is given by ${\mathcal C}_{\ssc p}=2 q_+ q_-$, 
and $q_+^2$ and $q_-^2$ are the eigenvalues of the corresponding
density matrix $\rho_{\!\ssc A\!B}$. Moreover, among the six parameters 
in $\left| \Psi  \rra$, $r$ uniquely characterizes the entanglement
and hence it is nonlocal. The phase parameter $\zeta$ is not exactly 
local as it does not belong to one of the two qubits, as the rest 
four. However, its value can be changed with a phase transformation 
on $\left| \phi_\perp  \rra$ or one on $\left| \phi'_\perp \rra$.
The $3\times 2=6$ parameter set of local $SU(2)$ transformations  
only affects the values of 5 out of the 6 parameters/coordinates 
in the description of the generic pure two-qubit state.

For a generic mixed state, one has a purification of it \cite{BZ}
in the sense that it can always be obtained as the reduced density
matrix of a pure state of a bipartite system. We start from the 
Schmidt decomposition for the latter given as
\[
\sum_{\ssc k=0}^{N-1} \sqrt{\mu_k} \left|e_k \rra \left|e'_k \rra \;,
\]
$0 \leq \sqrt{\mu_k} \leq 1$, $\sum_{\ssc k=0}^{N-1} \mu_k=1$, 
with $\left|e_k \rra$ and  $\left|e'_k \rra$ being orthonormal sets
of the two component Hilbert spaces having the same dimension $n$. 
To be definite, an ordering of the labeling should be specified,
and we generally take one such that the sequence of $\mu_k$ 
is a decreasing one.  We can express all $\left|e_k \rra$ and  
$\left|e'_k \rra$ for an arbitrary pure state in terms of unitary
transformations from a fixed basis. It is straightforward to
give a coordination of the pure state in terms of coordinates
for the $\left|e_k \rra$ and  $\left|e'_k \rra$. The above
expression for the two-qubit pure state of is one given by 
$\mu_{\ssc 0}=q_+^2$ and $\mu_{\ssc 1}=q_-^2$. Note that
we have to allow the relative phase $e^{i\zeta}$ as we have
arbitrarily fixed the phase in $\left| \phi_\perp \phi'_\perp \rra$.
Retrieving the mixed state as reduced density matrix of the 
purification gives us the coordination of the space of mixed 
state ${\mathcal{M}}$. The strategy works for any quNit system, 
{\em i.e.} system with an $N$-dimensional Hilbert space, though 
the number of coordinates of course grows fast with $N$.

Let us give an explicit description for such a
coordination/parametrization of qutrit states. From 
the Schmidt decomposition, we first express
 $\left|e_{\ssc 0} \rra$ as a unitary transform of
 $\left|0 \rra$, up to an overall phase factor. The little 
group is $SU(2)$ acting on the orthogonal subspace, and
we can take away an overall phase factor. Hence we need 
only $8-3-1=4$ parameters. The transform we denote 
by $U_{\!\ssc 20}$ given as the explicit matrix
\bea
U_{\!\ssc 20}= 
U_{\!\ssc 21} U_{\!\ssc 10} 
= \left( \begin{array}{cc}
1 & 0  \\
0 & \left( \begin{array}{cc}
c_{\!\ssc 21} & - \bar{s}_{\!\ssc 21}  \\
 s_{\!\ssc 21} & \bar{c}_{\!\ssc 21} \\
\end{array}  \right)
\end{array}  \right)  
  \left( \begin{array}{cc}
 \left( \begin{array}{cc}
c_{\!\ssc 10} & - \bar{s}_{\!\ssc 10} \\
s_{\!\ssc 10} & \bar{c}_{\!\ssc 10} \\
\end{array}  \right) & 0 \\
0 & 1
\end{array}  \right) \
=\left( \begin{array}{ccc}
c_{\!\ssc 10} & - \bar{s}_{\!\ssc 10} & 0 \\
c_{\!\ssc 21} s_{\!\ssc 10} & c_{\!\ssc 21} \bar{c}_{\!\ssc 10} &  - \bar{s}_{\!\ssc 21} \\
s_{\!\ssc 21} s_{\!\ssc 10}  & s_{\!\ssc 21} \bar{c}_{\!\ssc 10}  &  \bar{c}_{\!\ssc 21} 
\end{array}  \right) \;,
\eea
adopting the general relations of 
$c_{..} = e^{\frac{-i}{2}\psi_{..}} \cos{\frac{1}{2}\theta_{..}}$  
and $s_{..} = e^{\frac{i}{2}\psi_{..}} \sin{\frac{1}{2}\theta_{..}}$,
where  $_{..}$ represents any one of the composite indices.
Dropping all the phases, the column vector of the matrix,
and its higher-dimensional analogs, as a generic orthonormal 
basis of a real vector space coordinate has been widely used. 
The phases $\psi_{..}$ gives the relative phases between the
corresponding components. Generalization to arbitrary $N$
is obvious, {\em i.e.} one can take the fixed vector $\left|0 \rra$ 
to an arbitrary (normalized) one,  in an $n$-dimensional Hilbert,
up to an overall phase factor, space by a 
\bea
U_{\!\ssc (n-1)0} = U_{\!\ssc (n-1)(n-2)} U_{\!\ssc (n-2)(n-3)} \cdots U_{\!\ssc 10}
\eea
with $2(n-1)$ parameters. The little group is
 $SU(n-1)$. After writing $\left| e_{\ssc 0} \rra$ 
as $U_{\!\ssc 20} \left| 0 \rra$,  and similarly 
$\left| e'_{\ssc 0} \rra$ as $U'_{\!\ssc 20} \left| 0 \rra$,
we note that $\left| e_{\ssc 1} \rra$ and 
$\left| e_{\ssc 2} \rra$ live in a Hilbert space
spanned by  $U_{\!\ssc 20} \left| 1 \rra$,
and $U_{\!\ssc 20} \left| 2 \rra$. We can simply
repeat the strategy to write $\left| e_{\ssc 1} \rra$,
up to an overall phase factor, as 
$\tilde{U}_{\!\ssc 21} U_{\!\ssc 20}\!  \left| 1 \rra$
through a new $\tilde{U}_{\!\ssc 21}$. 
$\tilde{U}_{\!\ssc 21} U_{\!\ssc 20}\!  \left| 2\rra$
then has to be  $\left| e_{\ssc 2} \rra$ up to a phase
factor. Hence, a generic two-qutrit pure state can be written as
\[
\sqrt{\mu_{\ssc 0}} U_{\!\ssc 20}\!  \left| 0 \rra   \otimes U'_{\!\ssc 20}\!  \left| 0 \rra 
   +   \sqrt{\mu_{\ssc 1}}  e^{i\zeta_1} \tilde{U}_{\!\ssc 21} U_{\!\ssc 20}\!  \left| 1 \rra  
           \otimes \tilde{U}'_{\!\ssc 21} U'_{\!\ssc 20}\!  \left| 1 \rra
  +  \sqrt{\mu_{\ssc 2}}    e^{i\zeta_21}  \tilde{U}_{\!\ssc 21} U_{\!\ssc 20}\!  \left| 2 \rra  
       \otimes \tilde{U}'_{\!\ssc 21} U'_{\!\ssc 20}\!  \left| 2 \rra \;.
\]
where we have put in the relative phases $e^{i\zeta_1}$ and
$e^{i\zeta_2}$.  A word of caution in the reading of our naive 
operator notation of the unitary transformations is in order 
here. The $\tilde{U}_{\!\ssc 21}$, and similarly $\tilde{U}'_{\!\ssc 21}$, 
is a unitary transformation acting only within the vector subspace 
spanned by ${U}_{\!\ssc 20}\!  \left| 1\rra$ and 
$U_{\!\ssc 20}\!  \left| 2\rra$. While it has the general form 
of $SU(2)$ matrix adopted here on the basis of the two vectors, 
putting it in the form of the $U_{\!\ssc 21}$ as a $3\times 3$ 
matrix would be wrong. $\tilde{U}_{\!\ssc 21}$ does not 
act on $\left| e_{\ssc 0} \rra = U_{\!\ssc 20} \left| 0 \rra$
the coordinate vector of which is given by the first column 
of $U_{\!\ssc 20}$. (The right matrix product is to be taken 
in reverse order, in the original basis.)

To get back to single-qutrit mixed states, we simply have 
to take the reduced density matrix. We have
\[
 \mu_{\ssc 0}  U_{\!\ssc 20}  \left| 0 \rra\!\!\lla 0 \right| U_{\!\ssc 20}^\dag
+ \mu_{\ssc 1}  \tilde{U}_{\!\ssc 21} U_{\!\ssc 20}  \left| 1 \rra\!\!\lla 1 \right| U_{\!\ssc 20}^\dag \tilde{U}_{\!\ssc 21}^\dag
 +  \mu_{\ssc 2}  \tilde{U}_{\!\ssc 21} U_{\!\ssc 20}  \left| 2 \rra\!\!\lla 2 \right| U_{\!\ssc 20}^\dag \tilde{U}_{\!\ssc 21}^\dag
\]
as a generic qutrit state realized in terms of their eigenensembles.
The eigenvalues are exactly the $\mu_k$. In the kind of expressions as 
the above for the mixed state, one can think about the $U_{\!\ssc 20}$
and $\tilde{U}_{\!\ssc 21}$ type of symbols as referring to the operators
content of which as illustrated by the matrix expression for $U_{\!\ssc 20}$
given above. We have actually an expression for any mixed state, here for
a qutrit, in terms of the proper number of independent parameters which
can be seen as providing a coordinate system for the space of mixed states.
We have here eight parameters, the $3\times 2$ in $U_{\!\ssc 21}$,
$U_{\!\ssc 10}$ and $\tilde{U}_{\!\ssc 21}$ plus the two independent 
ones among $\mu_k$. A simple good choice (in general) is 
$1\geq \nu_i \equiv i(\mu_{i-{\ssc 1}} - \mu_i) \geq 0$, $i \ne 0$.
The explicit real coordinates, in this case, are given by 
$\theta_{\!\ssc 21}$, $\psi_{\!\ssc 21}$, $\theta_{\!\ssc 10}$, 
$\psi_{\!\ssc 10}$, $\tilde\theta_{\!\ssc 21}$, $\tilde\psi_{\!\ssc 21}$,
$\nu_{\ssc 1}$, and $\nu_{\ssc 2}$.  However, for many purposes,
using the $\mu_k$ set may be more convenient.

Similarly, one can give a generic mixed state for single qutetrait as
\bea&&
  \mu_{\ssc 0} U_{\!\ssc 30}  \left| 0 \rra\!\!\lla 0 \right| U_{\!\ssc 30}^\dag
+   \mu_{\ssc 1} \tilde{U}_{\!\ssc 31} U_{\!\ssc 30}  \left| 1 \rra\!\!\lla 1 \right| U_{\!\ssc 30}^\dag \tilde{U}_{\!\ssc 31}^\dag 
\sea \hspace*{.7in} 
 +   \mu_{\ssc 2} \widetilde{U}_{\!\ssc 32} \tilde{U}_{\!\ssc 31} U_{\!\ssc 30}  \left| 2 \rra\!\!\lla 2 \right| U_{\!\ssc 30}^\dag \tilde{U}_{\!\ssc 31}^\dag \widetilde{U}_{\!\ssc 32}^\dag 
 +  \mu_{\ssc 3} \widetilde{U}_{\!\ssc 32} \tilde{U}_{\!\ssc 31} U_{\!\ssc 30}  \left| 3 \rra\!\!\lla 3 \right| U_{\!\ssc 30}^\dag \tilde{U}_{\!\ssc 31}^\dag \widetilde{U}_{\!\ssc 32}^\dag \;.
\nonumber \eea
Note that $\widetilde{U}_{\!\ssc 32}$ is a different operator from 
$\tilde{U}_{\!\ssc 32}$ inside $\tilde{U}_{\!\ssc 31}=\tilde{U}_{\!\ssc 32}\tilde{U}_{\!\ssc 21}$.
One can easily extrapolate from here to the large $n$ cases and check that
the approach does give the right answer for the dimensions of the pure
state and mixed state spaces in general. 

\subsection{Modified Parametrization for a Bipartite System}
Next, we turn to a more interesting case, that of mixed two-qubit states.
A two-qubit system can, first of all, be seen as one of a single qutetrait.
From the result for the two-qutetrait pure state to single qutetrait mixed
 state reduction, we can first take the identification $\left| 0 \rra =\left| 00 \rra$, 
$\left| 1 \rra =\left| 11 \rra$, $\left| 2 \rra =\left| 01 \rra$, and
$\left| 3 \rra =\left| 10 \rra$, and replace $U_{\!\ssc 30} \left| 0 \rra$
by our earlier expression for a generic two-qubit pure state. It can indeed 
be easily appreciated that the latter can be seen as the result of a unitary
transformation acting on $\left| 00 \rra$. Explicitly, we have
$U_{\!\ssc 22}$ as an operator being given by 
$U_q (U_\phi \otimes U_{\phi'})$ where 
\bea
U_q= \left( \begin{array}{cccc}
 q_+ & -e^{-i\zeta} q_-  \\
e^{i\zeta} q_- &  q_+ 
\end{array}  \right)
\eea
acts in the vector subspace spanned by the basis vector 
$\left|\phi\phi' \rra$ and its orthogonal complement 
$\left|\phi_{\!\ssc\perp} \phi'_{\!\ssc\perp} \rra$. One can have 
$q_+=\cos{\chi}$, $q_-=\sin{\chi}$, with $\cos{2\chi}=r$, $\chi$ 
being the Schmidt angle of the mixing for the single-qubit state.
 The transformation $U_{\!\ssc 22}$ is to be adopted 
for the replacement of $U_{\!\ssc 30}$ in the expression 
above. The trick enforces the six coordinates for the description 
of a pure two-qubit state as part of the full set of fifteen coordinates 
for the description of a generic one. More specifically, we take our
four $\left| e_k \rra$ states as $U_{\!\ssc 22}  \left| 0 \rra$,
$U_{\!\ssc 31} U_{\!\ssc 22}  \left| 1 \rra$, 
$\tilde{U}_{\!\ssc 32} U_{\!\ssc 31} U_{\!\ssc 22}  \left| 2 \rra$,  
and $\tilde{U}_{\!\ssc 32} U_{\!\ssc 31} U_{\!\ssc 22}  \left| 3 \rra$, 
respectively, and the parallel for four $\left| e'_k \rra$ states.
Those parameters inside $U_{\!\ssc 31}$ and 
$\tilde{U}_{\!\ssc 32}$, and  the three $\nu_i$ gives  the extra 
nine coordinates needed to fully describe any mixed two-qubit state 
in the fifteen dimensional space.  Explicitly, we have 
\bea&&
\left| e_{\!\ssc 0} \rra = U_{\!\ssc 22}  \left| 0 \rra =   \left| \Psi \rra  
  = q_+ \left| \phi \phi' \rra  
    + e^{i\zeta} q_- \left| \phi_{\!\ssc\perp} \phi'_{\!\ssc\perp}  \rra
\sea
\left| e_{\!\ssc 1} \rra = {U}_{\!\ssc 31} U_{\!\ssc 22}  \left| 1 \rra
  = {U}_{\!\ssc 31} ( -e^{-i\zeta} q_- \left| \phi \phi' \rra  + q_+  \left| \phi_{\!\ssc\perp}  \phi'_{\!\ssc\perp} \rra )
\sea\hspace*{.3in}
 = c_{\!\ssc 21}  (-e^{-i\zeta} q_- \left| \phi \phi' \rra  
  +  q_+ \left| \phi_{\!\ssc\perp}  \phi'_{\!\ssc\perp} \rra )
   + s_{\!\ssc 21} ( c_{\!\ssc 32}  \left| \phi \phi'_{\!\ssc\perp}  \rra + s _{\!\ssc 32}   \left| \phi_{\!\ssc\perp}  \phi'\rra )
\sea
\left| e_{\!\ssc 2} \rra = \tilde{U}_{\!\ssc 32} U_{\!\ssc 31} U_{\!\ssc 22}  \left| 2 \rra 
 =  \tilde{U}_{\!\ssc 32} U_{\!\ssc 31} \left| \phi \phi'_{\!\ssc\perp}  \rra 
= {c}_{\!\ssc 0}  U_{\!\ssc 31} \left| \phi \phi'_{\!\ssc\perp}  \rra  + {s}_{\!\ssc 0} U_{\!\ssc 31}\left| \phi_{\!\ssc\perp}  \phi'\rra
\sea\hspace*{.3in}
 =   - {c}_{\!\ssc 0} \,\bar{s}_{\!\ssc 21} ( -e^{-i\zeta}  q_-  \left| \phi \phi' \rra  
  +   q_+   \left| \phi_{\!\ssc\perp}  \phi'_{\!\ssc\perp} \rra )
  + ( {c}_{\!\ssc 0} \bar{c}_{\!\ssc 21} c_{\!\ssc 32}   - {s}_{\!\ssc 0} \bar{s}_{\!\ssc 32}) \left| \phi \phi'_{\!\ssc\perp}  \rra + ( {c}_{\!\ssc 0}  \bar{c}_{\!\ssc 21} s_{\!\ssc 32} + {s}_{\!\ssc 0} \bar{c}_{\!\ssc 32} ) \left| \phi_{\!\ssc\perp}  \phi'\rra
\sea
\left| e_{\!\ssc 3} \rra = \tilde{U}_{\!\ssc 32} {U}_{\!\ssc 31} U_{\!\ssc 22}  \left| 3 \rra 
 =  \tilde{U}_{\!\ssc 32} U_{\!\ssc 31} \left| \phi_{\!\ssc\perp}  \phi'  \rra 
= - \bar{s}_{\!\ssc 0} {U}_{\!\ssc 31}  \left| \phi \phi'_{\!\ssc\perp}  \rra   + \bar{c}_{\!\ssc 0} {U}_{\!\ssc 31} \left| \phi_{\!\ssc\perp}  \phi'  \rra 
\sea\hspace*{.3in}
 =  \bar{s}_{\!\ssc 0} \,\bar{s}_{\!\ssc 21} (-e^{-i\zeta}  q_-  \left| \phi \phi' \rra  
  + q_+    \left| \phi_{\!\ssc\perp}  \phi'_{\!\ssc\perp} \rra )
  -( \bar{s}_{\!\ssc 0}  \bar{c}_{\!\ssc 21}  c_{\!\ssc 32} +\bar{c}_{\!\ssc 0}\, \bar{s}_{\!\ssc 32}) \left| \phi \phi'_{\!\ssc\perp}  \rra - (\bar{s}_{\!\ssc 0}  \bar{c}_{\!\ssc 21} s_{\!\ssc 32}- \bar{c}_{\!\ssc 0} \bar{c}_{\!\ssc 32} ) \left| \phi_{\!\ssc\perp}  \phi'\rra \!,
\sea\label{ek}
\eea
where we have used simply ${c}_{\!\ssc 0}$ and ${s}_{\!\ssc 0}$ 
for the parameter in $\tilde{U}_{\!\ssc 32}$. We may also use the
notation $\left| \Psi_{\!\ssc\perp} \rra \equiv {U}_{\!\ssc 22}   \left| 1 \rra$.
The expressions above is all we need to write down a generic two-qubit 
(mixed) state, and it shows the convenient use of orthonormal basis 
$\left| {\bf 0} \rra \equiv \left| \phi \phi' \rra$, 
$\left| {\bf 1} \rra \equiv \left| \phi_{\!\ssc\perp}  \phi'_{\!\ssc\perp} \rra$, 
$\left| {\bf 2} \rra \equiv \left| \phi  \phi'_{\!\ssc\perp} \rra$,
and $\left| {\bf 3} \rra \equiv \left| \phi_{\!\ssc\perp}  \phi' \rra$,
hiding the four local parameters in $\left| \phi \rra$ 
and $\left| \phi' \rra$. The effect of local, relative, phase 
transformations given by one on $\left| \phi_{\!\ssc\perp} \rra$ 
or $\left| \phi'_{\!\ssc\perp} \rra$ has effects on the  eleven 
parameters/coordinates explicitly shown in the expressions 
that takes a bit more effort to trace. We leave the full expression 
of the mixed two-qubit state density matrix to an appendix, for 
reference. We will refer to the basis as the bold-type basis below. It 
is a natural set of basis for the Hilbert state of the two-qubit pure states.

It is interesting to note that the scaled Hilbert-Schmidt
distance \cite{BZ} between an arbitrary state $\rho$ and 
$E_{\ssc\Psi}\equiv \left| \Psi \rra \!\!\lla  \Psi \right|$ 
is simply given by
\bea
D_{\!\ssc 2}(\rho, E_{\ssc\Psi})\equiv &&\frac{1}{\sqrt{2}} D_{\!\ssc H\!S}(\rho, E_{\ssc\Psi})
=\sqrt{\frac{1-2\mbox{Tr} (\rho E_{\ssc\Psi})+\mbox{Tr}\rho^2}{2}}
\sea\label{HSd}
  =\sqrt{(1- \mu_{\ssc 0})^2 - (\mu_{\ssc 1}\mu_{\ssc 2} +\mu_{\ssc 2}\mu_{\ssc 3}+ \mu_{\ssc 3}\mu_{\ssc 1})} \;.
\eea
It is easy to see $E_{\ssc\Psi}$ is really the pure state 
that is the closest to $\rho$. Moreover, 
$S_{\ssc\Psi}\equiv \left| {\bf 0} \rra\!\! \lla  {\bf 0} \right|$  
is the product state closest to the generally entangled pure state 
$E_{\ssc\Psi}$ at $D_{\!\ssc 2}(S_{\ssc\Psi}, E_{\ssc\Psi}) = q_-$ \cite{LS}.
These are the interesting features that are not available in 
the popular parametrization/coordination of ${\mathcal M}$.
Apart from giving  a clear and direct identification of the 
subspace of pure states 
(at $\nu_{\!\ssc 1}=1$, $\nu_{\!\ssc 2}=\nu_{\!\ssc 3}=0$) 
of dimension six, it allows direct implementation of important
entanglement analyses like the PPT condition, which can be
conveniently performed in the bold-type basis. 

What has been described above for the two-qubit case can again 
be extrapolated to other two-quNit cases. Of course with growth
of the dimension of the Hilbert space, the number of
parameters/coordinates required to describe all mixed
states may make it far too tedious to apply analytically.

\section{Some Entanglement Features and Interference of Entanglement}
\subsection{Entanglement Results and Relations for the Different Pure State Parts}
We are interested in applying our parametrization of two-qubit
mixed states to look at entanglement features. Actually, there is
something to learn from first looking at the pure states involved
in the eigenensemble expression. The concurrence, or equivalently
negativity, of a two-qubit pure state is essentially the sole measure
of its entanglement. We have for $E_{\ssc\Psi}$ and 
$E_{\ssc\Psi_{\!\ssc\perp}}$ the concurrence 
${\mathcal C}_{\ssc\Psi}= {\mathcal C}_{\ssc\Psi_{\!\ssc\perp}} (={\mathcal C}_p)$.  Consider 
$E_{e_1} \equiv \left| e_{\!\ssc 1} \rra\!\! \lla  e_{\!\ssc 1} \right|$.
The state vector is a linear combination of two orthonormal 
(generally) entangled states $\left| \Psi_{\!\ssc\perp} \rra$ and 
$\left| \Psi_{\!\ssc 1} \rra \equiv c_{\!\ssc 32}  \left| \phi \phi'_{\!\ssc\perp}  \rra 
     + s _{\!\ssc 32}   \left| \phi_{\!\ssc\perp}  \phi'\rra$, 
{\em cf.} Eq.(\ref{ek}), with concurrences given by 
${\mathcal C}_p =2q_+q_-$  and  ${\mathcal C}_{\ssc\Psi_{\!\ssc 1}} = 2 s _{\!\ssc 32} c_{\!\ssc 32}$,
respectively.  For $E_{e_1} $, we have the concurrence 
\bea
{\mathcal C}_{e_1}= | 2q_+ q_-  e^{i\zeta} \bar{c}_{\!\ssc 21}^2    +  2{c}_{\!\ssc 32}  {s}_{\!\ssc 32} \bar{s}_{\!\ssc 21}^2 |
  =  | {\mathcal C}_p  e^{i\zeta} \bar{c}_{\!\ssc 21}^2    + {\mathcal C}_{\ssc\Psi_{\!\ssc 1}} \bar{s}_{\!\ssc 21}^2 | \;,
\eea
showing an interesting pattern of interference of entanglement,
depending sensitively on the relative phases in the linear 
combination of the two parts. For fixed  ${\mathcal C}_p$, 
${\mathcal C}_{\ssc\Psi_1}$, and $\theta_{\!\ssc 21}$, ${\mathcal C}_{e_1}$ 
attains its maximum and minimum at  $e^{i(2\psi_{\ssc 21}+\zeta)} = \pm 1$, of
$|  {\mathcal C}_p \cos^2\!\!\left(\frac{\theta_{\ssc 21}}{2}\right)
   \pm {\mathcal C}_{\ssc\Psi_{\!\ssc 1}}  \sin^2\!\!\left(\frac{\theta_{\ssc 21}}{2}\right) \!|$,
respectively. The asymmetric appearance of $e^{i\zeta}$ 
here is an artifact of our parametrization. If one writes 
$\left| \Psi_{\!\ssc\perp} \rra$ as a linear combination 
of $\left| \phi \phi'  \rra$ and $\left| \phi_{\!\ssc\perp}  \phi'_{\!\ssc\perp}   \rra$
in the same form as $\left| \Psi_{\!\ssc 1}\rra$, interference 
depending only  on the relative phase between the two entangled 
parts would be obtained. Similarly, we can see the same  
behavior looking into ${\mathcal C}_{e_2}$. Explicitly, we have
\bea
{\mathcal C}_{e_2}=| {\mathcal C}_p e^{i\zeta} \bar{c}_{\!\ssc 0}^2 {s}_{\!\ssc 21}^2 
   +  2(\bar{c}_{\!\ssc 0} c_{\!\ssc 21} \bar{s}_{\!\ssc 32} + \bar{s}_{\!\ssc 0}  {c}_{\!\ssc 32})
     (\bar{c}_{\!\ssc 0} c_{\!\ssc 21} \bar{c}_{\!\ssc 32}  - \bar{s}_{\!\ssc 0}  {s}_{\!\ssc 32}) | \;,
\eea
where $|2(\bar{c}_{\!\ssc 0} c_{\!\ssc 21} \bar{s}_{\!\ssc 32}  + \bar{s}_{\!\ssc 0}  {c}_{\!\ssc 32})
 (\bar{c}_{\!\ssc 0} c_{\!\ssc 21} \bar{c}_{\!\ssc 32}  - \bar{s}_{\!\ssc 0}  {s}_{\!\ssc 32}) |$
is just $|{s}_{\!\ssc 0}|^2 +|{c}_{\!\ssc 0}|^2 |c_{\!\ssc 21}|^2$
times the concurrence of the normalized state vector $\left| \Psi_{\!\ssc 2} \rra$, 
as a linear combination of  $\left| \phi \phi'_{\!\ssc\perp}  \rra$ 
and $\left| \phi_{\!\ssc\perp}  \phi'\rra$,   {\em cf.} Eq.(\ref{ek}),  
and the normalization factor 
$\sqrt{|{s}_{\!\ssc 0}|^2 +|{c}_{\!\ssc 0}|^2 |c_{\!\ssc 21}|^2}$
is exactly the amplitude of the part of the normalized vector in
$\left| e_{\!\ssc 2} \rra$. The rest in the concurrence expression
is about the phases. Explicitly,
\bea&&
\left| \Psi_{\!\ssc 2} \rra =  \frac{1} { \sqrt{|{s}_{\!\ssc 0}|^2 +|{c}_{\!\ssc 0}|^2 |c_{\!\ssc 21}|^2}}  
 \left[  ( {c}_{\!\ssc 0} \bar{c}_{\!\ssc 21} c_{\!\ssc 32}   - {s}_{\!\ssc 0} \bar{s}_{\!\ssc 32}) \left| \phi \phi'_{\!\ssc\perp}  \rra + ( {c}_{\!\ssc 0}  \bar{c}_{\!\ssc 21} s_{\!\ssc 32} + {s}_{\!\ssc 0} \bar{c}_{\!\ssc 32} ) \left| \phi_{\!\ssc\perp}  \phi'\rra
 \right] \;,
\nonumber \\
&&  \hspace*{1.2in}
{\mathcal C}_{\ssc\Psi_{\!\ssc 2}} 
= \frac{2|(\bar{c}_{\!\ssc 0} c_{\!\ssc 21} \bar{s}_{\!\ssc 32}  + \bar{s}_{\!\ssc 0}  {c}_{\!\ssc 32})
 (\bar{c}_{\!\ssc 0} c_{\!\ssc 21} \bar{c}_{\!\ssc 32}  - \bar{s}_{\!\ssc 0}  {s}_{\!\ssc 32}) |}
   { |{s}_{\!\ssc 0}|^2 +|{c}_{\!\ssc 0}|^2 |c_{\!\ssc 21}|^2} \;, 
\eea
In exact parallel, we have 
\bea
 {\mathcal C}_{e_3} =|{\mathcal C}_p e^{i\zeta}  {s}_{\!\ssc 0}^2 {s}_{\!\ssc 21}^2
    + 2 ({s}_{\!\ssc 0} c_{\!\ssc 21} \bar{c}_{\!\ssc 32}  +{c}_{\!\ssc 0}  {s}_{\!\ssc 32} )
     (  {s}_{\!\ssc 0} c_{\!\ssc 21} \bar{s}_{\!\ssc 32}   - {c}_{\!\ssc 0}  {c}_{\!\ssc 32} )| \;,
\eea
with 
\bea &&
\left| \Psi_{\!\ssc 3} \rra =  \frac{1} { \sqrt{|{c}_{\!\ssc 0}|^2 +|{s}_{\!\ssc 0}|^2 |c_{\!\ssc 21}|^2}}
 \left[ -( \bar{s}_{\!\ssc 0} c_{\!\ssc 32} \bar{c}_{\!\ssc 21}  +\bar{c}_{\!\ssc 0}\, \bar{s}_{\!\ssc 32}) \left| \phi \phi'_{\!\ssc\perp}  \rra - (\bar{s}_{\!\ssc 0} s_{\!\ssc 32} \bar{c}_{\!\ssc 21} - \bar{c}_{\!\ssc 0} \bar{c}_{\!\ssc 32} ) \left| \phi_{\!\ssc\perp}  \phi'\rra \right] \;,
\nonumber \\
&&  \hspace*{1.2in}
 {\mathcal C}_{\ssc\Psi_{\!\ssc 3}} = \frac{2 |({s}_{\!\ssc 0} c_{\!\ssc 21} \bar{c}_{\!\ssc 32}  +{c}_{\!\ssc 0}  {s}_{\!\ssc 32} )
     (  {s}_{\!\ssc 0} c_{\!\ssc 21} \bar{s}_{\!\ssc 32}   - {c}_{\!\ssc 0}  {c}_{\!\ssc 32} )|}
   { |{c}_{\!\ssc 0}|^2 +|{s}_{\!\ssc 0}|^2 |c_{\!\ssc 21}|^2} \;.
\eea
Note that $\left| \Psi_{\!\ssc 1} \rra$, $\left| \Psi_{\!\ssc 2} \rra$,
and  $\left| \Psi_{\!\ssc 3} \rra$ are generally mutually not 
orthogonal, though they are orthogonal to $\left| \Psi\rra$ and
$\left| \Psi_{\!\ssc \perp} \rra$.

A mixed state as a particular mixture of pure states can be seen 
to have a concurrence as the weighted average of the concurrence
of the pure states in the mixture. However, a mixed state or
a density matrix can be written in different ways as a mixture.
The concurrence, as a general measure of entanglement, is to 
be taken as the convex roof or the minimum among all possible
mixture expressions. Hence, our results for the concurrence of
the four pure states $E_k \equiv \left| e_k \rra\!\! \lla e_k\right|$
provide at least an upper bound for the concurrence of any
mixed state of the two-qubit system, in terms of the eleven
parameters. Entanglement features are independent of the 
purely local parameters in the expressions of $\left|\phi \rra$ 
and  $\left|\phi' \rra$. However, we will see interesting cases
below where the entanglement within different parts of a mixture 
cancels against one another. That seems to be more the rule 
than the exception. Another measure of entanglement particularly 
useful for the two-qubit states is the negativity, as twice the magnitude 
of the single negative eigenvalue of the partial transpose. 
Recall that for pure states, there is only a single notion
of entanglement and the concurrence and negativity are equal. 
The difference between the two results for the mixed state is 
particularly interesting. Note that the partial transpose, as 
a Hermitian matrix, has a characteristic polynomial with only 
real coefficients. The complexity of the matrix elements, from our
parametrization, rests on the complex phases of four mixing
parameters of the pure state vector $\left| e_k \rra$ in terms 
of the states of the  bold-type basis. The parameters can only 
show up in the PPT condition and the negativity, and, we expect, 
also the concurrence, in specific real combinations.  We believe
we have the answer for them in our results for the entanglement
of the $\left| e_k \rra$ states, {\em i.e.} as the ${\mathcal C}_{e_1}$,
${\mathcal C}_{e_2}$, and ${\mathcal C}_{e_3}$ above. In fact,
it is likely that any entanglement results for the system can be
expressed in terms of the ten generally independent parameters
of $\nu_i$, ${\mathcal C}_{p}$, ${\mathcal C}_{\ssc\Psi_i}$,
and ${\mathcal C}_{e_i}$.

\subsection{Entanglement Results for a Large Class of Two-qubit States}
The entanglement feature of a special class of the generic mixed 
two-qubit states with a single of the fifteen parameters being zero is 
easy to analyze. The results are of course interesting in themselves.
We also see them as very illustrative of how the entanglement of 
the pure states in its eigenensemble work with, or more like against,
one another in contributing to the entanglement of the mixed state 
in general. One may compare that with the naive upper bound of the 
weighted average, which simply gives, for the concurrence for example,
${\mathcal C}_\rho \leq \sum_k {\mu_k} {\mathcal C}_{e_k}$
for the two-qubit states. When $\theta_{\!\ssc 21}=0$, {\em i.e.} 
${s}_{\!\ssc 21}=0$, the density matrix reduces to two $2\times 2$ 
diagonal blocks, a form that maintains in its partial transpose. The 
PPT condition is then given by the non-negativity of the determinants 
of the diagonal blocks of the latter, which is the two inequalities 
\bea
\mu_{\ssc 0 }  \mu_{\ssc 1} +  (\mu_{\ssc 0}-\mu_{\ssc 1})^2 q_+^2 q_-^2 \geq 
  (\mu_{\ssc 2}-\mu_{\ssc 3})^2 | {c}_{\!\ssc 0}\, \bar{c}_{\!\ssc 21}  {s}_{\!\ssc 32}  + s_{\!\ssc 0}\, \bar{c}_{\!\ssc 32}|^2 |\bar{c}_{\!\ssc 0}\,{c}_{\!\ssc 21}  \bar{c}_{\!\ssc 32}  -\bar{s}_{\!\ssc 0}{s}_{\!\ssc 32}|^2 
\eea
and
\bea
\mu_{\ssc 2}\mu_{\ssc 3}
+ (\mu_{\ssc 2}-\mu_{\ssc 3})^2  | {c}_{\!\ssc 0}\, \bar{c}_{\!\ssc 21}  {s}_{\!\ssc 32}  + s_{\!\ssc 0}\, \bar{c}_{\!\ssc 32}|^2 |\bar{c}_{\!\ssc 0}\,{c}_{\!\ssc 21}  \bar{c}_{\!\ssc 32}  -\bar{s}_{\!\ssc 0}{s}_{\!\ssc 32}|^2
\geq  (\mu_{\ssc 0}-\mu_{\ssc 1})^2 q_+^2 q_-^2  \;.
\eea
The first is trivially satisfied  as 
$\mu_{\ssc 0 }  \mu_{\ssc 1} \geq  (\mu_{\ssc 2}-\mu_{\ssc 3})^2$.
The other inequality is a true condition, which is equivalent to
\bea
4 \mu_{\ssc 2}\mu_{\ssc 3}
+ (\mu_{\ssc 2}-\mu_{\ssc 3})^2 {\mathcal C}_{\ssc\Psi_{\!\ssc 2}}^2  
\geq  (\mu_{\ssc 0}-\mu_{\ssc 1})^2 {\mathcal C}_p^2  \;,
\label{ppt-s}
\eea
where we have used $|{s}_{\!\ssc 0}|^2 +|{c}_{\!\ssc 0}|^2 |c_{\!\ssc 21}|^2=1$ 
as $|c_{\!\ssc 21}|=1$. Note that for the case, $\left| e_{\!\ssc 2} \rra$
and $\left| e_{\!\ssc 3} \rra$ reduce to two orthonormal linear
combinations of  $\left| \phi \phi'_{\!\ssc\perp}  \rra$ 
and $\left| \phi_{\!\ssc\perp}  \phi'\rra$, hence 
${\mathcal C}_{e_2}={\mathcal C}_{\ssc\Psi_{\!\ssc 2}}={\mathcal C}_{\ssc\Psi_{\!\ssc 3}}={\mathcal C}_{e_3}$; 
also we have ${\mathcal C}_{e_1}={\mathcal C}_{e_0}={\mathcal C}_p$. 
Considering particularly the case of $\mu_{\ssc 2}=\mu_{\ssc 3}=0$, it 
gives $(\mu_{\ssc 0}-\mu_{\ssc 1}) {\mathcal C}_p$ as the negativity of 
the resulting density matrix. That measures the entanglement of the state 
$\mu_{\ssc 0} E_{\!\ssc\Psi} + \mu_{\ssc 1} E_{\!\ssc\Psi_{\!\perp}}$,
or that part of the full state $\rho$.  This is a direct contrast to the
weighted sum of  $(\mu_{\ssc 0}+\mu_{\ssc 1}) {\mathcal C}_p$. 
Similarly, one can see $(\mu_{\ssc 2}-\mu_{\ssc 3}) {\mathcal C}_{\ssc\Psi_{\!\ssc 2}}$
as negativity of the other part, namely 
$\mu_{\ssc 2} E_{e_2} + \mu_{\ssc 3} E_{e_3} 
  =\mu_{\ssc 2} E_{\!\ssc\Psi_2} + \mu_{\ssc 3} E_{\!\ssc\Psi_3}$.
Directly checking the negativity of the general ($s_{\!\ssc 21}=0$) case, 
we get, for the entangled states violating the above inequality,
\bea\label{n-s}
{\mathcal N}_{s_{21}=0} = \sqrt{(\mu_{\ssc 0}-\mu_{\ssc 1})^2 {\mathcal C}_p^2
   + ({\mu_{\ssc 2}-\mu_{\ssc 3}})^2 (1-{\mathcal C}_{\ssc\Psi_{\!\ssc 2}}^2) }
-({\mu_{\ssc 2}+\mu_{\ssc 3}}) \;.
\eea
This shows the interesting but complicated interplay between 
the entanglement of the parts, as illustrated by the values of 
the negativities (which is the same as the concurrences for the 
pure states), of the two parts here of a mixed state contributed 
to the full state.  Calculating the concurrence is only a little bit 
more involved \cite{W}. The result reads
\begin{table}[t]
\begin{center}
\begin{tabular}{||c|c|c|c|c|c|c|c|c||}    \hline\hline
& 	$E_{\ssc\Psi}=E_{e_0}$
& 	$E_{\ssc\Psi_{\!\ssc 1}}$	& 	$E_{\ssc\Psi_{\!\ssc 2}}$	& 	$E_{\ssc\Psi_{\!\ssc 3}}$	& 	$E_{e_1}$	&	$E_{e_2}$	&	$E_{e_3}$ & PPT condition for $\rho=\sum \mu_k E_{e_k}$
  \\ \hline 
${s}_{\!\ssc 21}=0$  	&	${\mathcal C}_p$	&	${\mathcal C}_{\ssc\Psi_{\!\ssc 1}}$	& 	${\mathcal C}_{\ssc\Psi_{\!\ssc 2}}$	& 	${\mathcal C}_{\ssc\Psi_{\!\ssc 2}}$	&	${\mathcal C}_p$	& 	${\mathcal C}_{\ssc\Psi_{\!\ssc 2}}$	& 	${\mathcal C}_{\ssc\Psi_{\!\ssc 2}}$	
	&	$4 \mu_{\ssc 2}\mu_{\ssc 3}+ (\mu_{\ssc 2}-\mu_{\ssc 3})^2 {\mathcal C}_{\ssc\Psi_{\!\ssc 2}}^2  \geq  (\mu_{\ssc 0}-\mu_{\ssc 1})^2 {\mathcal C}_p^2$		   \\
\hline 
${c}_{\!\ssc 21}=0$  	&	${\mathcal C}_p$	&	${\mathcal C}_{\ssc\Psi_{\!\ssc 1}}$	& 	${\mathcal C}_{\ssc\Psi_{\!\ssc 1}}$	& 	${\mathcal C}_{\ssc\Psi_{\!\ssc 1}}$	& 	${\mathcal C}_{\ssc\Psi_{\!\ssc 1}}$	& 	I(${\mathcal C}_p$,${\mathcal C}_{\ssc\Psi_{\!\ssc 1}}$)	& 	I(${\mathcal C}_p$,${\mathcal C}_{\ssc\Psi_{\!\ssc 1}}$)		
	& $\mu_{\ssc 2}=\mu_{\ssc 3}$:  $4 \mu_{\ssc 0}\mu_{\ssc 2}+ (\mu_{\ssc 0}-\mu_{\ssc 2})^2 {\mathcal C}_p^2
\geq  (\mu_{\ssc 1}-\mu_{\ssc 2})^2 {\mathcal C}_{\ssc\Psi_{\!\ssc 1}}^2$ \\
&&&&&&&&
 \ \ \ and \  \ \ $4 \mu_{\ssc 1}\mu_{\ssc 2}+ (\mu_{\ssc 1}-\mu_{\ssc 2})^2 {\mathcal C}_{\ssc\Psi_{\!\ssc 1}}^2  
\geq  (\mu_{\ssc 0}-\mu_{\ssc 2})^2 {\mathcal C}_p^2$\\
\hline 
${s}_{\!\ssc 32}=0$  	&	${\mathcal C}_p$	&	0									& 	${\mathcal C}_{\ssc\Psi_{\!\ssc 2}}'$	& 	${\mathcal C}_{\ssc\Psi_{\!\ssc 3}}'$	& 	$|{c}_{\!\ssc 21}|^2{\mathcal C}_p$		& 	I(${\mathcal C}_p$,${\mathcal C}_{\ssc\Psi_{\!\ssc 2}}'$)		& 	I(${\mathcal C}_p$,${\mathcal C}_{\ssc\Psi_{\!\ssc 3}}'$)		&		${\mathcal C}_p=0$,  ($\mu_{\ssc 2}=\mu_{\ssc 3}$ or ${c}_{\!\ssc 21}=0$)  : satisfied\\
\hline 
${c}_{\!\ssc 32}=0$  	&	${\mathcal C}_p$	&	0									& 	${\mathcal C}_{\ssc\Psi_{\!\ssc 2}}'$	& 	${\mathcal C}_{\ssc\Psi_{\!\ssc 3}}'$	& 	$|{c}_{\!\ssc 21}|^2{\mathcal C}_p$		& 	I(${\mathcal C}_p$,${\mathcal C}_{\ssc\Psi_{\!\ssc 2}}'$)		& 	I(${\mathcal C}_p$,${\mathcal C}_{\ssc\Psi_{\!\ssc 3}}'$)			&    ${\mathcal C}_p=0$,  ($\mu_{\ssc 2}=\mu_{\ssc 3}$ or ${c}_{\!\ssc 21}=0$) : satisfied	\\
\hline 
${s}_{\!\ssc 0}=0$		&	${\mathcal C}_p$	&	${\mathcal C}_{\ssc\Psi_{\!\ssc 1}}$	&	${\mathcal C}_{\ssc\Psi_{\!\ssc 1}}$	&	${\mathcal C}_{\ssc\Psi_{\!\ssc 1}}$	&	I(${\mathcal C}_p$,${\mathcal C}_{\ssc\Psi_{\!\ssc 1}}$)		&	I(${\mathcal C}_p$,${\mathcal C}_{\ssc\Psi_{\!\ssc 1}}$)		&	${\mathcal C}_{\ssc\Psi_{\!\ssc 1}}$	
  	 & $\mu_{\ssc 1}=\mu_{\ssc 2}$:  
$4 \mu_{\ssc 1}\mu_{\ssc 3}+ (\mu_{\ssc 1}-\mu_{\ssc 3})^2 {\mathcal C}_{\ssc\Psi_{\!\ssc 1}}^2  
\geq  (\mu_{\ssc 0}-\mu_{\ssc 1})^2 {\mathcal C}_p^2$\\
&&&&&&&&
${c}_{\!\ssc 21}=0$ :    $4 \mu_{\ssc 0}\mu_{\ssc 2}+ (\mu_{\ssc 0}-\mu_{\ssc 2})^2 {\mathcal C}_p^2
\geq  (\mu_{\ssc 1}-\mu_{\ssc 3})^2 {\mathcal C}_{\ssc\Psi_{\!\ssc 1}}^2$ \\
&&&&&&&&
 \ \ \ and \  \ \ $4 \mu_{\ssc 1}\mu_{\ssc 3}+ (\mu_{\ssc 1}-\mu_{\ssc 3})^2 {\mathcal C}_{\ssc\Psi_{\!\ssc 1}}^2  
\geq  (\mu_{\ssc 0}-\mu_{\ssc 2})^2 {\mathcal C}_p^2$\\
\hline 
${c}_{\!\ssc 0}=0$		&	${\mathcal C}_p$	&	${\mathcal C}_{\ssc\Psi_{\!\ssc 1}}$	&	${\mathcal C}_{\ssc\Psi_{\!\ssc 1}}$	&	${\mathcal C}_{\ssc\Psi_{\!\ssc 1}}$	& 	I(${\mathcal C}_p$,${\mathcal C}_{\ssc\Psi_{\!\ssc 1}}$)		&	${\mathcal C}_{\ssc\Psi_{\!\ssc 1}}$	&	I(${\mathcal C}_p$,${\mathcal C}_{\ssc\Psi_{\!\ssc 1}}$)		
 &	 ${c}_{\!\ssc 21}=0$ :    $4 \mu_{\ssc 0}\mu_{\ssc 3}+ (\mu_{\ssc 0}-\mu_{\ssc 3})^2 {\mathcal C}_p^2
\geq  (\mu_{\ssc 1}-\mu_{\ssc 2})^2 {\mathcal C}_{\ssc\Psi_{\!\ssc 1}}^2$ \\
&&&&&&&&
 \ \ \ and \  \ \ $4 \mu_{\ssc 1}\mu_{\ssc 2}+ (\mu_{\ssc 1}-\mu_{\ssc 2})^2 {\mathcal C}_{\ssc\Psi_{\!\ssc 1}}^2  
\geq  (\mu_{\ssc 0}-\mu_{\ssc 3})^2 {\mathcal C}_p^2$\\
    \hline\hline
\end{tabular}
\caption{Entanglement of the pure states at the various limits 
and some related entanglement results for the mixed state :
The I$(\cdot,\cdot)$ symbol denotes an entanglement as the interference of the two parts,
{\em i.e.} as the magnitude of a normalized complex linear combination of the two
 entanglement measures. Note that in general, 
${\mathcal C}_{e_i}= \mbox{I}({\mathcal C}_p$,${\mathcal C}_{\ssc\Psi_{\!\ssc i}})$. 
Recall that
${\mathcal C}_p =2q_+q_-$ and ${\mathcal C}_{\ssc\Psi_{\!\ssc 1}} = 2 s _{\ssc 32} c_{\ssc 32}$.
We  have also    ${\mathcal C}_{\ssc\Psi_{\!\ssc 2}}'
 =\frac{2|{c}_{\ssc 0}{s}_{\ssc 0} {c}_{\ssc 21}|}{|{s}_{\ssc 0}|^2+ |{c}_{\ssc 0}|^2 |{c}_{\ssc 21}|^2}$
and   ${\mathcal C}_{\ssc\Psi_{\!\ssc 3}}'
 =\frac{2|{c}_{\ssc 0}{s}_{\ssc 0} {c}_{\ssc 21}|}{|{c}_{\ssc 0}|^2+ |{s}_{\ssc 0}|^2 |{c}_{\ssc 21}|^2}$.
In each of the above cases, the concurrence is given by the square root of the right-hand side 
of the PPT inequality minus that of the left-hand side, and the negativity is given by 
 $\sqrt{ f({\mathcal C}_p^2,{\mathcal C}_{\ssc\Psi_{\!\ssc 1}}^2) + (\mu_h -\mu_k)^2} -  (\mu_h +\mu_k)$ 
with the PPT condition taken in the form 
$4\mu_h \mu_k  \geq  f({\mathcal C}_p^2,{\mathcal C}_{\ssc\Psi_{\!\ssc 1}}^2)$
($\mu_h \geq \mu_k$, {\em i.e.} $h<k$),
when the PPT condition is violated.
}
\hrulefill
\end{center}
\end{table}
\bea\label{c-s}
{\mathcal C}_{s_{21}=0} = (\mu_{\ssc 0}-\mu_{\ssc 1}) {\mathcal C}_p
   - \sqrt{ ({\mu_{\ssc 2}-\mu_{\ssc 3}})^2 {\mathcal C}_{\ssc\Psi_{\!\ssc 2}}^2 
                   + 4 \mu_{\ssc 2} \mu_{\ssc 3}} \;.
\eea
We see that $(\mu_{\ssc 0}-\mu_{\ssc 1}) {\mathcal C}_p$ is 
also the concurrence for the state 
$\mu_{\ssc 0} E_{\!\ssc\Psi} + \mu_{\ssc 1} E_{\!\ssc\Psi_{\!\perp}}$.
That suggests that the negativity and concurrence of a mixed state
with an eigenensemble as weighted sum of two exact orthogonal
complements of the same two basis pure states, as vectors of the
Hilbert space, has the same value as the difference between the 
two parts.  We note that the concurrence equals to the negativity 
and they are independent of ${\mathcal C}_{\ssc\Psi_{\!\ssc 2}}$ 
when $\mu_{\ssc 2}=\mu_{\ssc 3}$.  Another interesting special 
limit to check is when $\mu_{\ssc 3}$ vanishes. For the concurrence, 
we have the simple result of $(\mu_{\ssc 0}-\mu_{\ssc 1}) {\mathcal C}_p
            - \mu_{\ssc 2} {\mathcal C}_{\ssc\Psi_{\!\ssc 2}}$. 
That of the negativity, explicitly as
 $\sqrt{(\mu_{\ssc 0}-\mu_{\ssc 1})^2 {\mathcal C}_p^2
    - \mu_{\ssc 2}^2 {\mathcal C}_{\ssc\Psi_{\!\ssc 2}}^2
   + \mu_{\ssc 2}^2  }-\mu_{\ssc 2}$, 
is a bit more complicated, as the $\mu_k$ parameters are not
completely independent. In general, both the negativity and the 
concurrence results are increasing functions of all, the independent,
 $\nu_i$ parameters. And they have essentially the maximum value
of $\mu_{\ssc 0} {\mathcal C}_p$ which can attain unity.

In the $s_{\!\ssc 21}=0$ case, we basically have the entanglement
as having the $(\mu_{\ssc 0}-\mu_{\ssc 1}) {\mathcal C}_p$ from
the dominating $\mu_{\ssc 0} E_{e_0} + \mu_{\ssc 1} E_{e_1}$
contribution, which in itself is the difference between the two 
parts, being further partially canceled by that of the
$\mu_{\ssc 2} E_{e_2} + \mu_{\ssc 3} E_{e_3}$ part.  Short of 
having the generic entanglement results for the two-qubit state,
we look at the other limits to check plausible similar features
of entanglement cancellation. In Table~1, the entanglement
results for the various pure states involved are given at the
various limits of the mixing parameters inside the pure states.
One can see that the $c_{\!\ssc 21}=0$, ${c}_{\!\ssc 0}=0$, or  
${s}_{\!\ssc 0}=0$ cases all involve only the two basic pure state
entanglement as ${\mathcal C}_p$ and ${\mathcal C}_{\ssc\Psi_{\!\ssc 1}}$.
With an extra condition imposed, for example $\mu_{\ssc 2}=\mu_{\ssc 3}$
for the $c_{\!\ssc 21}=0$ case, we have explicit results of some
interest. In particular, for the last case, the PPT condition is given by
\bea\label{ppt-c23'}
4 \mu_{\ssc 0}\mu_{\ssc 2}
+ (\mu_{\ssc 0}-\mu_{\ssc 2})^2 {\mathcal C}_p^2
\geq  (\mu_{\ssc 1}-\mu_{\ssc 2})^2 {\mathcal C}_{\ssc\Psi_{\!\ssc 1}}^2    \;,
\eea
and
\bea\label{ppt-c23}
4 \mu_{\ssc 1}\mu_{\ssc 2}
+ (\mu_{\ssc 1}-\mu_{\ssc 2})^2 {\mathcal C}_{\ssc\Psi_{\!\ssc 1}}^2  
\geq  (\mu_{\ssc 0}-\mu_{\ssc 2})^2 {\mathcal C}_p^2  \;.
\eea
The negativity is
\bea&&
{\mathcal N}_{c_{21}=0, \mu_{\ssc 2}=\mu_{\ssc 3}} 
= \sqrt{ ({\mu_{\ssc 1}-\mu_{\ssc 2}})^2 {\mathcal C}_{\ssc\Psi_{\!\ssc 1}}^2 
+ (\mu_{\ssc 0}-\mu_{\ssc 2})^2 (1-{\mathcal C}_p^2) }
-({\mu_{\ssc 0}+\mu_{\ssc 2}}) \;,
\sea\label{n-c23}
\mbox{or} \quad
{\mathcal N}_{c_{21}=0, \mu_{\ssc 2}=\mu_{\ssc 3}} 
= \sqrt{(\mu_{\ssc 0}-\mu_{\ssc 2})^2 {\mathcal C}_p^2
   + ({\mu_{\ssc 1}-\mu_{\ssc 2}})^2 (1-{\mathcal C}_{\ssc\Psi_{\!\ssc 1}}^2) }
-({\mu_{\ssc 1}+\mu_{\ssc 2}}) \;,
\eea
corresponding to the cases of the first and second inequalities 
above being violated, respectively. The matching concurrence
is, respectively,
\bea &&
{\mathcal C}_{c_{21}=0, \mu_{\ssc 2}=\mu_{\ssc 3}}
= (\mu_{\ssc 1}-\mu_{\ssc 2}){\mathcal C}_{\ssc\Psi_{\!\ssc 1}}   - \sqrt{ ({\mu_{\ssc 0}-\mu_{\ssc 2}})^2 {\mathcal C}_p^2    + 4 \mu_{\ssc 0} \mu_{\ssc 2}} \;,
\sea\label{c-c23}
\mbox{or} \quad
{\mathcal C}_{c_{21}=0, \mu_{\ssc 2}=\mu_{\ssc 3}} 
=  (\mu_{\ssc 0}-\mu_{\ssc 2}){\mathcal C}_p   - \sqrt{ ({\mu_{\ssc 1}-\mu_{\ssc 2}})^2 {\mathcal C}_{\ssc\Psi_{\!\ssc 1}}^2    + 4 \mu_{\ssc 1} \mu_{\ssc 2}} \;.
\eea
Here, depending on which entanglement of the key two parts 
dominate, {\em i.e.} if $(\mu_{\ssc 0}-\mu_{\ssc 2}){\mathcal C}_p$ 
is smaller or larger than 
$(\mu_{\ssc 1}-\mu_{\ssc 2}){\mathcal C}_{\ssc\Psi_{\!\ssc 1}}$,
it is only when the difference of the two parts is large
enough that we have an entanglement that increases with
${\mathcal C}_{\ssc\Psi_{\!\ssc 1}}$ and decreases with 
${\mathcal C}_p$ or  increases with ${\mathcal C}_p$
and decreases with ${\mathcal C}_{\ssc\Psi_{\!\ssc 1}}$,
respectively.  That applies to both the negativity and the
concurrence. The $(\mu_{\ssc 0}-\mu_{\ssc 2}){\mathcal C}_p$
dominating case is an exact parallel with the $s_{\!\ssc 21}=0$ 
case above. When $(\mu_{\ssc 1}-\mu_{\ssc 2}){\mathcal C}_{\ssc\Psi_{\!\ssc 1}}$
dominates, however, it is somewhat different. While the
concurrence has the form 
$\mu_{\ssc 1} {\mathcal C}_{\ssc\Psi_{\!\ssc 1}} -\mu_{\ssc 0} {\mathcal C}_p$
for $\mu_{\ssc 2}=0$,  and generally can still attain 
$\mu_{\ssc 1} {\mathcal C}_{\ssc\Psi_{\!\ssc 1}}$, 
its value is bounded by $\frac{1}{2}$ as the maximum 
$\mu_{\ssc 1}$ from the $\mu_k$ hierarchy. The negativity
has maximum in the form 
$\sqrt{ \mu_{\ssc 1}^2 {\mathcal C}_{\ssc\Psi_{\!\ssc 1}}^2 +\mu_{\ssc 0}^2} -\mu_{\ssc 0}$
which, while still decreases with an increase in $\mu_{\ssc 0}$,
can only get to $\frac{\sqrt{2}-1}{2}$ at the maximum
$\mu_{\ssc 0}$ of $\frac{1}{2}$.

Other similar cases we have explicit entanglement results for
are given in Table~1. We refrain from presenting an explicit
analysis of their basic features, which are much of a parallel 
to those discussed above.

\subsection{Applications to Generalizations of Werner States}

In terms of the scaled Hilbert-Schmidt geometry, the maximally
mixed state ($\nu_i=0$) and the lines connecting it at the center 
to the four corners of a tetrahedral (taken as the pure states
$\left| e_k \rra\!\! \lla e_k\right|$ here) is particularly interesting
 \cite{BZ}. These are generalized Werner states \cite{We,LS}. 
We write a generalized Werner state for our system as 
$x \left| \Phi \rra\!\! \lla \Phi \right| + \frac{1-x}{4} I$, 
$x\in [-\frac{1}{3}, 1]$, where $\left| \Phi \rra$ gives a pure 
state with any, not necessarily maximal, entanglement. We first 
take the case of $\left| \Phi \rra$ being $\left| e_{\ssc 0} \rra$, 
{\em i.e.} $\left| \Psi \rra$. We are really looking at the subspace 
${\mathcal{M}}_{\ssc (13)}$ in Ref.\cite{BZ}, with 
$\mu \equiv \mu_{\ssc 1} = \mu_{\ssc 2} =\mu_{\ssc 3} \;(\leq \frac{1}{4})$, 
hence $\mu_{\ssc 0} = 1-3 \mu $. Any such mixed state clearly 
is described by the seven parameters/coordinates, including the 
six for the pure state of Eq.(\ref{2qps}) and the mixing parameter 
$\mu$, or $\nu_{\ssc 1} (= 1-4\mu)$ which is the $x$ in this case,
except that no negative value of it is allowed. Explicitly, we have 
the line $W_{\!\mu} \equiv (1-4\mu) E_{\ssc\Psi}  + \mu I$.
We want to check the separability of an even bigger set of states, 
one which has the same form of the density matrix in the bold-type
basis but with no constraint on the $\mu_k$. We denote that by
 $W_{\!\vec\mu}$, which can be obtained by imposing 
$\theta_{\!\ssc 32}= \theta_{\!\ssc 21}= \theta_{\!\ssc 0}=0$, 
hence $s_{\!\ssc 32}= s_{\!\ssc 21}= s_{\!\ssc 0}=0$. 
Explicitly, $W_{\!\vec\mu}= \mu_{\ssc 0} E_{\!\ssc\Psi} + \mu_{\ssc 1} E_{\!\ssc\Psi_{\!\perp}}
+ \mu_{\ssc 2} \left| \phi \phi'_{\!\ssc\perp} \rra \!\!\lla \phi \phi'_{\!\ssc\perp} \right|
  + \mu_{\ssc 3} \left| \phi_{\!\ssc\perp} \phi' \rra\! \!\lla \phi_{\!\ssc\perp} \phi' \right|$.
It is only a special case of the $s_{\!\ssc 21}=0$ class we analyzed
in the previous section. The separability condition is given by 
${\mathcal C}_{p} \leq \frac{2\sqrt{\mu_{\ssc 2}\mu_{\ssc 3}}}{\mu_{\ssc 0}-\mu_{\ssc 1}}$,
which reduces to ${\mathcal C}_{p} \leq \frac{2\mu}{1-4\mu}= \frac{1-\nu_{\ssc 1}}{2\nu_{\ssc 1}}$ 
for the generalized Werner states $W_{\!\mu}$. In fact, for the
inseparable cases,  we have the negativity 
\bea
 {\mathcal N}_{W_{\!\vec\mu}} 
 =  \sqrt{ {\mathcal C}_{p}^2 (\mu_{\ssc 0}-\mu_{\ssc 1})^2 + (\mu_{\ssc 2}-\mu_{\ssc 3})^2}
  -  (\mu_{\ssc 2}+\mu_{\ssc 3}) \;,
\eea
and concurrence
\bea
{\mathcal C}_{W_{\!\vec\mu}}  = (\mu_{\ssc 0}-\mu_{\ssc 1}) {\mathcal C}_p
   - 2\sqrt{  \mu_{\ssc 2} \mu_{\ssc 3}} \;,
\eea
hence ${\mathcal N}_{W_{\!\mu}} = {\mathcal C}_{W_{\!\mu}} 
 = (1-4\mu) {\mathcal C}_{p} -2\mu
=\nu_{\ssc 1} {\mathcal C}_{p} -\frac{1-\nu_{\ssc 1}}{2}$.

The separability condition for $W_{\!\mu}$ can be rewritten as 
$[2(2{\mathcal C}_p-1) \mu -{\mathcal C}_p][2(2{\mathcal C}_p+1) \mu -{\mathcal C}_p]  \leq 0$,
which reduces to  $\mu \geq  \frac{{\mathcal C}_p}{2(2{\mathcal C}_p+1)}$, or 
$\nu_{\ssc 1} \leq  \frac{1}{2{\mathcal C}_p+1}$.  Note that we have, 
for $\nu_{\ssc 1} >  \frac{1}{2{\mathcal C}_p+1}$,
$ {\mathcal N}_{W_{\!\mu}} = \frac{(2{\mathcal C}_p+1) \nu_{\ssc 1} -1}{2}$.
In particular, we have  $\mu \geq  \frac{1}{6}$ and 
$\nu_{\ssc 1} \leq \frac{1}{3}$ for ${\mathcal C}_{p}=1$, and  $\mu \geq  \frac{1}{8}$ 
and $\nu_{\ssc 1} \leq \frac{1}{2}$ for ${\mathcal C}_{p}=\frac{1}{2}$. 
For the limit of small ${\mathcal C}_{p}$, the separability condition
tends to $\mu \geq  \frac{{\mathcal C}_p}{2}$, or $\nu_{\ssc 1} \leq 1-2{\mathcal C}_p$
linear in ${\mathcal C}_p$. Of course for ${\mathcal C}_{p}=0$, all of $W_{\!\mu}$ are 
separable. One can easily obtain $D_{\ssc 2} (W_{\!\mu}, E_{\ssc\Psi})=\sqrt{6}\mu$.  
The result shows  that one can consider the $W_{\!\mu}$ set 
as a straight line through $E_{\ssc\Psi}$ with $\mu$ as effectively
a measure of distance \cite{LS}, we have established clearly for 
any ${\mathcal C}_p$ value. The separable $W_{\!\mu}$ state closest to 
$E_{\ssc\Psi}$ is given by $\mu = \frac{{\mathcal C}_p}{2(2{\mathcal C}_p+1)}$.\footnote{
The product state $S_{\ssc\Psi}$ as the particular $E_{\ssc\Psi}$
at $r=1$ is the product state that is closest to $E_{\ssc\Psi}$, as 
shown in Ref.\cite{LS}, at a distance of $\sqrt{\frac{1-r}{2}}$. The 
maximally mixed state $W_{\!\frac{1}{4}}$ is at an $r$-independent 
distance of  $\frac{\sqrt{6}}{4}$ from any $E_{\ssc\Psi}$, including 
$S_{\ssc\Psi}$. From simple trigonometry, the point on $W_{\!\mu}$ 
at the projection of $S_{\ssc\Psi}$ is at a distance of $\frac{1-r}{\sqrt{6}}$
from $E_{\ssc\Psi}$ and has a $\mu$ value of $\frac{1-r}{6}$, giving 
$\frac{2\mu}{(1-4\mu)}=\frac{1-r}{1+2r}$. But ${\mathcal C}_p^2=1-r^2 \geq (1-r)^2$ 
giving ${\mathcal C}_p  \geq \frac{1-r}{1+2r}$ with equality only for 
$r=0$. Hence, the points give a state that is entangled except for the 
case of $r=0$. The result directly contradicts a statement in the key 
`theorem' of  Ref.\cite{LS} the proof of which we fail to appreciate. 
Even taking only that maximally entangled as $E_{\ssc\Psi} (r=0)$ 
to get the $W_{\!\mu} (r=0)$ as a baseline does not work. While 
the perpendicular from $S_{\ssc\Psi}$ to $W_{\!\mu} (r=0)$ is 
right at the separable boundary, for another $W_{\!\mu} (r)$ from 
an $E_{\ssc\Psi}(r)$ between $S_{\ssc\Psi}$ and the maximally 
entangled one, {\em i.e.} with generic $r \in (0,1)$, the intersecting 
point of that $W_{\!\mu} (r)$ and the perpendicular from 
$S_{\ssc\Psi}$ to $W_{\!\mu} (r=0)$ can be seen not to give 
a right distance to $W_{\!\frac{1}{4}}$ [along $W_{\!\mu} (r)$]
to agree with our separability results. A result like that of  
Ref.\cite{LS} could otherwise give a good base to obtain the
full explicit characterization of the separability under our
parametrization of the generic two-qubit state.
}
One can also check that 
$D_{\ssc 2} (W_{\!\mu}, W_{\!\frac{1}{4}})=\frac{\sqrt{6}}{4}\nu_{\ssc 1}$,
and $\nu_{\ssc 1}$ effectively measures the distance of $W_{\!\mu}$ 
from the maximally mixed state $W_{\!\frac{1}{4}}$.

The generalized Werner states for $x \in [-\frac{1}{3}, 0)$ can be 
simply analyzed by relaxing our condition of $\mu_k$ ordering.
It is like taking the same analysis with an inverted ordering 
for the original two-qutetrait pure state construction, starting with
$\sum_{\ssc k=0}^{n-1} \sqrt{\mu_{3-k}} \left|e_k \rra \!\left|e'_k \rra$.
It can easily be obtained that all such states are separable. It is not 
difficult to find the description of such states within our coordinate 
system. They are the ${\mathcal{M}}_{\ssc (31)}$ subspace with
$\mu' \equiv \mu_{\ssc 0} = \mu_{\ssc 1} =\mu_{\ssc 2}$, 
$\mu_{\ssc 3} = 1-3 \mu'$, with $\frac{1}{3}\geq \mu' \geq \frac{1}{4}$.
The only nonzero $\nu_i$ is $\nu_{\ssc 3} = 3(4 \mu' -1)$,
with the parameter $x$ as $1-4 \mu'$ having the exact 
admissible range of value. Explicitly, we have
$W_{\!\mu'} = (1-4\mu') E_{e_{\ssc 3}}  + \mu' I$ where
$E_{e_{\ssc 3}} \equiv \left|e_{\ssc 3} \rra\!\!\lla e_{\ssc 3}\right|$.
The corresponding separability condition for $C'_{p}$ being
concurrence of the pure state $\left|e_{\ssc 3} \rra$ is
 $C'_{p}\leq \frac{2\mu'}{4\mu'-1}$ trivially satisfied.

Actually, the above result for the separability of the 
$W_{\!\mu}$ and states implies that for any 
$x \left| \Phi \rra\!\! \lla \Phi \right| + \frac{1-x}{4} I$
leading from the maximally mixed state at the center to an 
arbitrary pure state $E_{\!\ssc \Phi}$ on the sphere of radius 
$R_{\mbox{\tiny out}}=\frac{\sqrt{6}}{4}$ where the pure 
states reside, the separable states extended from the center 
to a distance $\frac{\sqrt{6}}{4}x$ of  the value 
$\frac{\sqrt{6}}{4(2{\mathcal C}_p+1)}$ as constrained by the 
concurrence ${\mathcal C}_p$ value of $E_{\!\ssc \Phi}$ as the
pure state closest to the set of generalized Werner states. 

\section{Final Remarks}

We want to emphasize that our parametrization, for a generic
system of mixed states as well as the modified form for a bipartite
system, introduced is particularly interesting. As full coordination
of the space of mixed states, many interesting quantities and 
geometric features of the space \cite{BZ} have quite direct 
descriptions under the parameters. That is mostly the result 
of having the $\nu_i$, or hierarchical $\mu_k$, incorporated 
as part of the parameters/coordinates, giving the unambiguous 
identification of the subspaces of the pure states with the 
geometric result of Eq.{(\ref{HSd}). This is especially so for 
the case of the bipartite systems. For the two-qubit case, for 
example, the parametrization identifies the entanglement
as independent of the four, or generally $4(N-1)$ for two-quNit,
purely local parameters. Moreover, one has entanglements 
of the pure state parts in the eigenensemble, {\em i.e.} the
 ${\mathcal C}_{p}$, ${\mathcal C}_{\ssc\Psi_i}$, and 
${\mathcal C}_{e_i}$, systematically identified. The intriguing
plausibility of expressing entanglement for any mixed state 
in terms of them together $\nu_i$ is worth further studies.
 
While the analytically technical difficulty to get to a general
parameterized complete description of any particular measure
of entanglement for all possible mixed states is still a hurdle 
to surmount even for the two-qubit system, not to say a more 
complicated one, our parametrization scheme may have merits 
for the studies of entanglement features in various parts of the 
space of all such mixed states. Our study here focuses only on 
the two-qubit case, but many qualitative features of our results are
expected to be somewhat applicable in a more general case. The
role of the entanglement of different parts, of a pure or mixed 
state, in contributing to the entanglement results of the whole,
we believe, is of interest both theoretically and experimentally. 
Our scheme for the parametrization may serve this kind of 
purpose well and similar analyses along the approach can be 
applied to other bipartite system with comparable results.

For two-qubit entanglement in particular, let us say a few words
about the difficulty in getting fully complete results. With
the parametrization of the general density matrix and hence its 
partial transpose, putting down the PPT condition is straightforward 
but still tedious and not very illustrative. For the partial transpose 
to be semi-positive, it requires all its principal minors to be 
nonnegative. That is $15-4=11$ inequalities. Even if the condition
of a nonnegative determinant suffices, for example in the case of
$c_{\!\ssc 21}=0$, reading physics out of the inequality is quite
nontrivial. To calculate the negativity or the concurrence is
about solving the corresponding quartic characteristic polynomial.
We fail to obtain a full analytical result. However, the latter may
not be an impossible task. Moreover, some semi-analytic or
numerical methods may be applied to help improve our
understanding of the (more) general results. That is not only
about appreciating the entanglement patterns of the mixed 
states. Because the parametrization is particularly well suited 
to illustrate the geometric structure of the full space of mixed
states, the results may help to clarify further the relationship 
between the two, as some of our results do. 

In looking at how the entanglement of the parts contribute to
that of the whole state, we have essentially illustrated a good
approach to identify such parts, and presented the results for
the whole as interference or partial cancellation between those
of the parts, for pure and nontrivial mixed states, respectively.
In the interest of our general parametrization scheme, we have
the eigenensemble mixing parameter $\mu_k$ restricted to have 
the hierarchy. For a more general application looking into some
specific classes of states of a certain system, being the two-qubit
one or otherwise, the basic approach and some of the results may
be used interestingly without that hierarchy imposed. 

\bigskip\bigskip

{\bf Appendix: The full two-qubit density matrix with our parametrization. }\\
--- the bold-typed basis : $\left| {\bf 0}\rra \equiv \left| \phi \phi' \rra$
 $\left| {\bf 1}\rra \equiv \left| \phi_{\!\ssc\perp} \phi'_{\!\ssc\perp} \rra$, 
$\left| {\bf 2}\rra \equiv \left| \phi \phi'_{\!\ssc\perp} \rra$, $\left| {\bf 3}\rra \equiv \left| \phi_{\!\ssc\perp} \phi' \rra$.
\bea &&
\rho_{\ssc\bf 00} : 
	(\mu_{\ssc 0}-\mu_{\ssc 3}) q_+^2 + (\mu_{\ssc 1}-\mu_{\ssc 3})  q_-^2 |c_{\!\ssc 21}|^2 + (\mu_{\ssc 2}-\mu_{\ssc 3})  q_-^2 |{c}_{\!\ssc 0}|^2 |s_{\!\ssc 21}|^2 + \mu_{\ssc 3}
\sea
\rho_{\ssc \bf 01}: 
	(\mu_{\ssc 0}-\mu_{\ssc 3}) e^{-i\zeta} q_- q_+ - (\mu_{\ssc 1}-\mu_{\ssc 3})  e^{-i\zeta} q_- q_+ |c_{\!\ssc 21}|^2 - (\mu_{\ssc 2}-\mu_{\ssc 3}) e^{-i\zeta} q_- q_+ |{c}_{\!\ssc 0}|^2 |s_{\!\ssc 21}|^2 
\sea
\rho_{\ssc\bf 02}: 
	-(\mu_{\ssc 1}-\mu_{\ssc 3})  e^{-i\zeta} q_- c_{\!\ssc 21} \bar{c}_{\!\ssc 32} \bar{s}_{\!\ssc 21}   + (\mu_{\ssc 2}-\mu_{\ssc 3})  e^{-i\zeta} q_-  {c}_{\!\ssc 0} \bar{s}_{\!\ssc 21}  (\bar{c}_{\!\ssc 0} c_{\!\ssc 21} \bar{c}_{\!\ssc 32}   - \bar{s}_{\!\ssc 0}  {s}_{\!\ssc 32} ) 
\sea
\rho_{\ssc\bf 03}:
	-(\mu_{\ssc 1}-\mu_{\ssc 3})  e^{-i\zeta} q_- c_{\!\ssc 21} \bar{s}_{\!\ssc 32} \bar{s}_{\!\ssc 21}   +(\mu_{\ssc 2}-\mu_{\ssc 3})  e^{-i\zeta} q_- {c}_{\!\ssc 0} \bar{s}_{\!\ssc 21} (  \bar{c}_{\!\ssc 0} c_{\!\ssc 21} \bar{s}_{\!\ssc 32}   + \bar{s}_{\!\ssc 0}  {c}_{\!\ssc 32} ) 
\sea 
\rho_{\ssc \bf 10}: 
	(\mu_{\ssc 0}-\mu_{\ssc 3})  e^{i\zeta} q_- q_+  - (\mu_{\ssc 1}-\mu_{\ssc 3})  e^{i\zeta} q_- q_+ |c_{\!\ssc 21}|^2  -(\mu_{\ssc 2}-\mu_{\ssc 3})  e^{i\zeta} q_- q_+ |{c}_{\!\ssc 0}|^2 |s_{\!\ssc 21}|^2 
\sea
\rho_{\ssc \bf 11}: 
	(\mu_{\ssc 0}-\mu_{\ssc 3}) q_-^2  + (\mu_{\ssc 1}-\mu_{\ssc 3}) q_+^2 |c_{\!\ssc 21}|^2 + (\mu_{\ssc 2}-\mu_{\ssc 3}) q_+^2 |{c}_{\!\ssc 0}|^2 |s_{\!\ssc 21}|^2  + \mu_{\ssc 3} 
\sea
\rho_{\ssc \bf 12}: 
	(\mu_{\ssc 1}-\mu_{\ssc 3})   q_+ c_{\!\ssc 21} \bar{c}_{\!\ssc 32} \bar{s}_{\!\ssc 21}  -(\mu_{\ssc 2}-\mu_{\ssc 3})   q_+ {c}_{\!\ssc 0} \bar{s}_{\!\ssc 21}  (\bar{c}_{\!\ssc 0} c_{\!\ssc 21} \bar{c}_{\!\ssc 32}  - \bar{s}_{\!\ssc 0}  {s}_{\!\ssc 32} )
\sea
\rho_{\ssc  \bf 13}: 
	(\mu_{\ssc 1}-\mu_{\ssc 3}) q_+ c_{\!\ssc 21} \bar{s}_{\!\ssc 32} \bar{s}_{\!\ssc 21}  -(\mu_{\ssc 2}-\mu_{\ssc 3})  q_+{c}_{\!\ssc 0} \bar{s}_{\!\ssc 21}  (\bar{c}_{\!\ssc 0} c_{\!\ssc 21} \bar{s}_{\!\ssc 32}    + \bar{s}_{\!\ssc 0}  {c}_{\!\ssc 32} )
\sea  
\rho_{\ssc \bf 20}: 
	-(\mu_{\ssc 1}-\mu_{\ssc 3}) e^{i\zeta} q_- \bar{c}_{\!\ssc 21} c_{\!\ssc 32} {s}_{\!\ssc 21}  +(\mu_{\ssc 2}-\mu_{\ssc 3}) e^{i\zeta} q_-  \bar{c}_{\!\ssc 0} {s}_{\!\ssc 21} ({c}_{\!\ssc 0} \bar{c}_{\!\ssc 21} {c}_{\!\ssc 32}   - {s}_{\!\ssc 0}\bar{s}_{\!\ssc 32} )
\sea
\rho_{\ssc \bf 21}: 
	(\mu_{\ssc 1}-\mu_{\ssc 3}) q_+\bar{c}_{\!\ssc 21} c_{\!\ssc 32} {s}_{\!\ssc 21}  -(\mu_{\ssc 2}-\mu_{\ssc 3})  q_+ \bar{c}_{\!\ssc 0} {s}_{\!\ssc 21} ({c}_{\!\ssc 0} \bar{c}_{\!\ssc 21} {c}_{\!\ssc 32}     - {s}_{\!\ssc 0}\bar{s}_{\!\ssc 32} )
\sea
\rho_{\ssc \bf 22}: 
	(\mu_{\ssc 1}-\mu_{\ssc 3}) |c_{\!\ssc 32}|^2 |s_{\!\ssc 21}|^2 +(\mu_{\ssc 2}-\mu_{\ssc 3}) 
| {c}_{\!\ssc 0}\, \bar{c}_{\!\ssc 21}  {c}_{\!\ssc 32}  - s_{\!\ssc 0}\, \bar{s}_{\!\ssc 32}|^2
+\mu_{\ssc 3}
\sea
\rho_{\ssc \bf 23}: 
(\mu_{\ssc 1} -\mu_{\ssc 3})  |s_{\!\ssc 21}|^2  {c}_{\!\ssc 32} \bar{s}_{\!\ssc 32} +
 (\mu_{\ssc 2}-\mu_{\ssc 3}) ( \bar{c}_{\!\ssc 0}\, {c}_{\!\ssc 21}  \bar{s}_{\!\ssc 32}  + \bar{s}_{\!\ssc 0}\, {c}_{\!\ssc 32})({c}_{\!\ssc 0}\,\bar{c}_{\!\ssc 21}  {c}_{\!\ssc 32}  -{s}_{\!\ssc 0} \bar{s}_{\!\ssc 32})
\sea 
\rho_{\ssc \bf 30}: 
	- (\mu_{\ssc 1} -\mu_{\ssc 3}) e^{i\zeta} q_- \bar{c}_{\!\ssc 21} s_{\!\ssc 32} {s}_{\!\ssc 21}  +(\mu_{\ssc 2} -\mu_{\ssc 3})  e^{i\zeta} q_-  \bar{c}_{\!\ssc 0} {s}_{\!\ssc 21} ( {c}_{\!\ssc 0} \bar{c}_{\!\ssc 21} {s}_{\!\ssc 32}   +{s}_{\!\ssc 0} \bar{c}_{\!\ssc 32} )
\sea
\rho_{\ssc \bf 31}: 
	(\mu_{\ssc 1} -\mu_{\ssc 3})  q_+\bar{c}_{\!\ssc 21} s_{\!\ssc 32} {s}_{\!\ssc 21}    -(\mu_{\ssc 2} -\mu_{\ssc 3})  q_+ \bar{c}_{\!\ssc 0} {s}_{\!\ssc 21}  ({c}_{\!\ssc 0} \bar{c}_{\!\ssc 21} {s}_{\!\ssc 32}    + {s}_{\!\ssc 0} \bar{c}_{\!\ssc 32})   
\sea
\rho_{\ssc \bf 32}: 
	(\mu_{\ssc 1} -\mu_{\ssc 3}) |s_{\!\ssc 21}|^2    \bar{c}_{\!\ssc 32} {s}_{\!\ssc 32}  
    +(\mu_{\ssc 2}-\mu_{\ssc 3}) ( {c}_{\!\ssc 0}\, \bar{c}_{\!\ssc 21}  {s}_{\!\ssc 32}  + s_{\!\ssc 0}\, \bar{c}_{\!\ssc 32})(\bar{c}_{\!\ssc 0}\,{c}_{\!\ssc 21}  \bar{c}_{\!\ssc 32}  -\bar{s}_{\!\ssc 0}{s}_{\!\ssc 32})
\sea
\rho_{\ssc \bf 33}: 
 (\mu_{\ssc 1} -\mu_{\ssc 3} ) |s_{\!\ssc 21}|^2 |s_{\!\ssc 32}|^2   +(\mu_{\ssc 2}-\mu_{\ssc 3}) | {c}_{\!\ssc 0}\, \bar{c}_{\!\ssc 21}  {s}_{\!\ssc 32}  + s_{\!\ssc 0}\, \bar{c}_{\!\ssc 32}|^2   +\mu_{\ssc 3}  
\nonumber\eea

\section*{Acknowledgments :}
The work is supported by the research grants number
109-2112-M-008-016 and 110-2112-M-008-016 of the MOST of Taiwan.\\

\noindent
{\bf Data availability :} the theoretical research reported involves no data.

 \end{document}